\documentclass[fleqn, usenatbib, useAMS]{mnras}
\usepackage{soul, hyperref, orcidlink}
\usepackage{enumitem, color, graphicx}
\usepackage{fleqn, times, amsmath, amssymb, amsfonts, latexsym,  esint}
\usepackage{array, mathtools, subfigure, esint, CJK}
\usepackage{makecell, multirow, ulem, bm}

\def\hmpc{\, {h {\rm Mpc}^{-1}}}           \def\mpch{\, {h^{-1} {\rm Mpc}}}

\def\dd{{\rm d}}                    
        
\def\vu{{\bf u}}          
\def\vx{{\bf x}}          

\title[Turbulent motion for cosmic fluid]{Turbulence, Thermal Pressure, and Their Dynamical Effects on Cosmic Baryonic Fluid}

\author[Y. Wang et al. ]{Yun Wang\orcidlink{0000-0003-4064-417X}$^1$\thanks{E-mail: yunw@jlu.edu.cn} and Ping He\orcidlink{0000-0001-7767-6154}$^{1,2}$\thanks{E-mail: hep@jlu.edu.cn} \\
	$^{1}$College of Physics, Jilin University, Changchun 130012, China. \\
	$^{2}$Center for High Energy Physics, Peking University, Beijing 100871, China.}

\date{Accepted XXX. Received YYY; in original form ZZZ}

\begin{document}
\maketitle 

\begin{abstract}
We employ the IllustrisTNG simulation data to investigate the turbulent and thermal motions of the cosmic baryonic fluid. With continuous wavelet transform techniques, we define the pressure spectra, or density-weighted velocity power spectra, as well as the spectral ratios, for both turbulent and thermal motions. We find that the magnitude of the turbulent pressure spectrum grows slightly from $z=4$ to $2$ and increases significantly from $z=2$ to $1$ at large scales, suggesting progressive turbulence injection into the cosmic fluid, whereas from $z=1$ to $0$, the spectrum remains nearly constant, indicating that turbulence may be balanced by energy transfer and dissipation. The magnitude of the turbulent pressure spectra also increases with environmental density, with the highest density regions showing a turbulent pressure up to six times that of thermal pressure. We also explore the dynamical effects of turbulence and thermal motions, discovering that while thermal pressure provides support against structure collapse, turbulent pressure almost counteracts this support, challenging the common belief that turbulent pressure supports gas against overcooling.
\end{abstract}

\section{Introduction}
\label{sec:intro}

The turbulent motion of the cosmic baryonic fluid in large-scale structures of the Universe has attracted increasing attention in cosmological studies over the last several decades \cite[e.g.][]{Bonazzola1987, Schuecker2004, Dolag2005, Hep2006, Schmidt2010, Zhu2010, Zhu2011a, Iapichino2011, Gaspari2013, Zhu2013, Fusco-Femiano2014, Shi2014, Bruggen2015, Vazza2018, Valdarnini2019, Angelinelli2020}. The existence of turbulence within the intracluster medium (ICM) has been confirmed by various observations. Direct evidence comes from the detection of non-thermal line broadening in X-ray emissions by the Hitomi satellite \citep{Hitomi2016}. Indirect evidence includes magnetic field fluctuations observed in the diffuse radio emissions from galaxy clusters \citep{Vogt2003, Murgia2004, Vogt2005, Ensslin2006, Bonafede2010, Vacca2010, Vacca2012}, X-ray surface brightness and pressure fluctuations derived from X-ray and Sunyaev-Zel'dovich effect maps \citep{Schuecker2004, Churazov2012, Gaspari2014, Zhuravleva2014, Walker2015, Khatri2016, Zhuravleva2018}, and the suppression of resonant scattering in X-ray spectra \citep{Churazov2004, Zhuravleva2013, Hitomi2018, Shi2019}.

The physical origin of turbulence in the cosmic baryonic fluid mainly consists of the following. Accretion through hierarchical mergers \citep{Subramanian2006, Bauer2012, Iapichino2017}, the associated physical processes, such as the injection and amplification of shock wave-induced vorticity \citep[e.g.][]{Ryu2008, Porter2015, Vazza2017} or ram pressure stripping \citep[e.g.][]{Cassano2005, Subramanian2006, Roediger2007}, should be able to generate and maintain turbulence in the intergalactic medium (IGM) and in galaxy clusters. Additionally, outflows or feedbacks from active galactic nucleus jets, and the supernova-driven galactic winds, are also expected to generate turbulence around galaxies and in cluster cores \citep{Bruggen2005, Iapichino2008b, Evoli2011, Gaspari2011, Iapichino2013, Banerjee2014, Angelinelli2020}.

Turbulent pressure is believed to contribute to the cosmic fluid's pressure support, a concept that can be traced back to the works of \citet{Chandra1951a, Chandra1951b}. This effect is reported to enhance the effective pressure support by approximately $10-30$\% in the cosmic baryonic fluid \citep{Lau2009, Eckert2019, Ettori2019, Angelinelli2020, Simonte2022}, mitigating the overcooling problem \citep{Fangtt2009, Zhu2010, Valdarnini2011, Schmidt2017}. The turbulence is also expected to contribute to the total pressure of the ICM, biasing the hydrostatic mass reconstruction \citep{Rasia2004, Lau2009, Morandi2011, Shi2015, Shi2016, Fusco-Femiano2018, Ota2018, Fusco-Femiano2019}.

In a previous work \citep{Wang2024b}, we use continuous wavelet transform (CWT) techniques to construct the global and environment-dependent wavelet statistics, such as energy spectrum and kurtosis, to study the fluctuation and intermittency of the turbulent motion in the cosmic fluid velocity field with the IllustrisTNG (TNG hereafter) simulation data. In this work, we continue to apply the CWT techniques to the TNG data to study the turbulent and thermal motions of the cosmic baryonic fluid in the large-scale structures of the Universe. Using the spectral ratio of turbulent to thermal pressure, we demonstrate that turbulent pressure dominates at smaller scales.  However, the dynamical investigations show that turbulent pressure often counteracts the support provided by thermal pressure, and thus cannot help prevent the overcooling problem.

\begin{figure*}
\centerline{\includegraphics[width=0.975\textwidth]{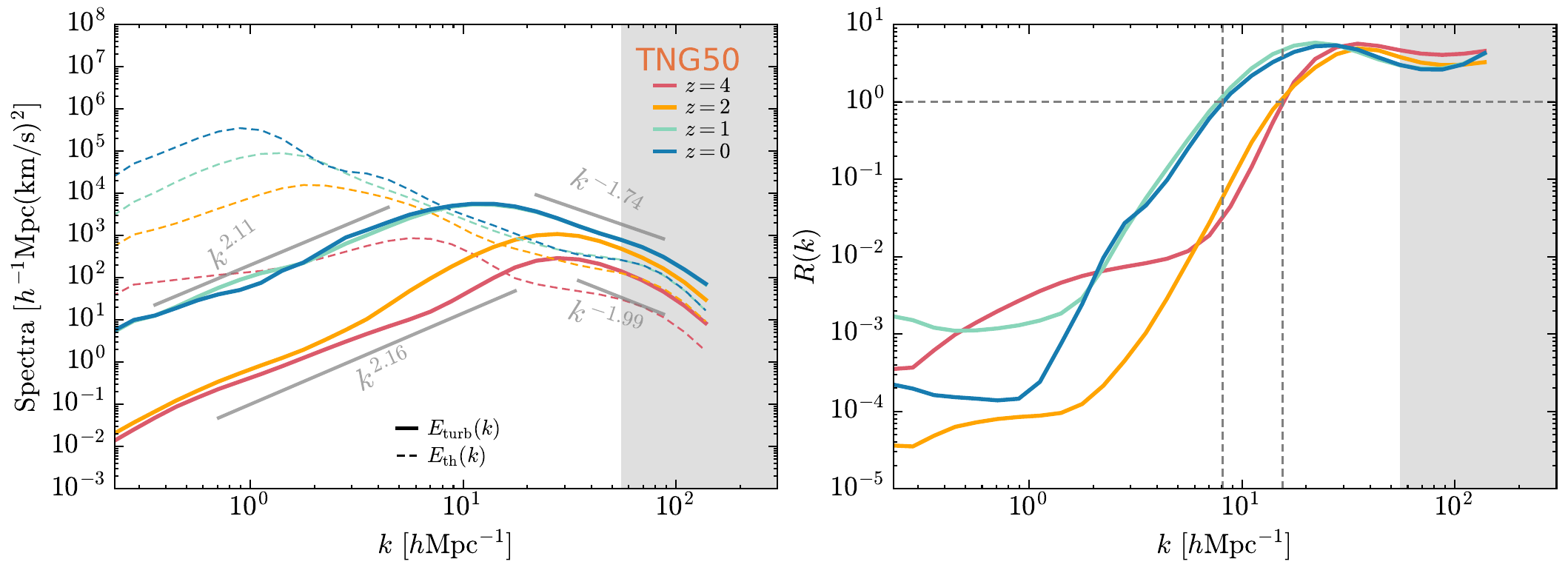}}
\caption{The $z$-evolution of the global energy spectrum  (left panel) and the spectral ratio (right panel) for the turbulent and thermal motions in the TNG50 simulation. The global energy spectrum $E(k)$ is obtained from the global WPS in Equation~(\ref{eq:global-wps}) as $E(k) = k^2 S(k)$. Notice there are peaks in all the energy spectra. The relevant power-law fits for the scale ranges that are both smaller and larger than the peak positions are indicated in the figure.} The vertical dashed lines indicate $k_{\rm equal}$, where $R(k_{\rm equal}) = 1$. The grey area shows the $k > 0.4 k_{\rm Nyquist}$ region.
\label{fig:TNG50_ES_global_evolution}
\end{figure*}

\section{Thermal and Turbulent Pressure}
\label{sec:pressure}

\subsection{Simulation Data}
\label{sec:data}

Throughout the work, we use the TNG50-1 simulation sample, corresponding to a simulated Universe of the size $35\mpch$, from the TNG project \citep{Pillepich2018a, Springel2018, Marinacci2018, Nelson2018, Naiman2018, Nelson2019}. TNG uses the moving-mesh code AREPO, which solves the cosmological gravity-magnetohydrodynamic equations on a dynamically unstructured mesh using a second-order accurate Godunov-type scheme \citep{Springel2010}, and \citet{Bauer2012} demonstrates its superiority in accurately modelling turbulence. Besides gravitational and hydrodynamic calculations, the TNG simulations also include a comprehensive set of physical processes, such as stellar formation and evolution, the associated metal enrichment, gas cooling, feedback from stellar wind, supernova and active galactic nucleus, and the magnetic field in cosmic structures \citep{Pillepich2018b, Nelson2019}. Given the abundance of the physical processes included in the TNG simulations and AREPO's excellent performance in hydrodynamical computations, the TNG data are well-suited for studying cosmic fluid turbulence.

\subsection{Pressure Ratio and Spectrum}
\label{sec:ratio-spec}

TNG provides the internal energy per unit mass of baryonic gas, $e_{\rm u}(\vx)$. Hence we have the thermal energy density,
\begin{flalign}
\label{eq:thermo-energy}
\varepsilon_{\rm th}(\vx) = \rho_{\rm b}(\vx) e_{\rm u}(\vx) = \frac{3}{2}\frac{\rho_{\rm b}(\vx)}{\mu(\vx) m_{\rm p}} k_{\rm B}T(\vx),
\end{flalign}
where $k_{\rm B}$ is the Boltzmann constant, and $m_{\rm p}$ the mass of the proton. $\mu(\vx)$, $\rho_{\rm b}(\vx)$ and $T(\vx)$, dependent on the spatial locations, are the mean molecular weight, mass density and temperature of the baryonic fluid, respectively. Regarding the cosmic fluid as the monoatomic ideal gas, its thermal pressure is
\begin{flalign}
\label{eq:thermo-pressure}
P_{\rm th}(\vx) = \frac{\rho_{\rm b}(\vx)}{\mu(\vx) m_{\rm p}} k_{\rm B}T(\vx) = \frac{2}{3}\varepsilon_{\rm th}(\vx).
\end{flalign}
The turbulent energy density is
\begin{flalign}
\label{eq:turb-energy}
\varepsilon_{\rm turb} = \frac{1}{2}\rho_{\rm b}(\vx) \vu^2_{\rm turb}(\vx),
\end{flalign}
in which $\vu_{\rm turb}(\vx)$ is the three-dimensional velocity of turbulent flow. In general, a velocity field of the fluid $\vu$ is a superposition of the turbulent velocity and the bulk velocity, i.e. $\vu = \vu_{\rm turb} + \vu_{\rm bulk}$. As in \citet{Wang2024b}, we also use the iterative multi-scale filtering approach developed by \citet{Vazza2012} to extract turbulent motions from the velocity field of the cosmic fluid. To use the bulk flow removal code in the Appendix of \citet{Vazza2012}, we set the relevant parameters as \texttt{eps=0.05}, \texttt{epssk=0.5} and \texttt{nk=16}, with the weighting factor $w_i=1$.

Consistent with the definition of thermal pressure in Equation~(\ref{eq:thermo-pressure}), we define the turbulent pressure as
\begin{flalign}
\label{eq:turb-pressure}
P_{\rm turb}(\vx)  \equiv \frac{2}{3}\varepsilon_{\rm turb}(\vx) = \frac{1}{3}\rho_{\rm b}(\vx) \vu^2_{\rm turb}(\vx).
\end{flalign}
The definition of turbulent pressure is consistent with that of \citet{Iapichino2008b, Lau2009, Paul2011, Angelinelli2020}. We also define the ratio of turbulent pressure to thermal pressure as
\begin{flalign}
\label{eq:pressure-ratio}
r(\vx) \equiv \frac{P_{\rm turb}(\vx)}{P_{\rm th}}  = \frac{\varepsilon_{\rm turb}(\vx)} {\varepsilon_{\rm th}(\vx)}= \frac{\vu^2_{\rm turb}(\vx)}{2e_{\rm u}(\vx)}.
\end{flalign}
For the random velocity field of the cosmic baryonic fluid $\vu_{\rm turb}(\vx)$ with the zero mean value, its isotropic CWT $\tilde{\vu}_{\rm turb}(w,\vx)$ is obtained by convolution with the wavelet function $\Psi$ as 
\begin{flalign}
\label{eq:wvt}
\tilde{\vu}_{\rm turb}(w,\vx) = \int\vu_{\rm turb}(\boldsymbol{\tau})\Psi(w, \vx - \boldsymbol{\tau})\dd^3\boldsymbol{\tau}.
\end{flalign}
Throughout this work, we use the so-called 3D isotropic cosine-weighted Gaussian-derived wavelet (CW-GDW), which can achieve good localization in both spatial and frequency space simultaneously \citep{Wang2022b, Wang2024a, Wang2024b}. With $\tilde{\vu}_{\rm turb}(w,\vx)$, and referring to Equation~(\ref{eq:turb-energy}), we define the local wavelet power spectrum (WPS) for the turbulent velocity $\tilde{\vu}_{\rm turb}(w,\vx)$ as
\begin{flalign}
\label{eq:spect-turb}
S_{\rm turb}(w, \vx) \equiv \frac{1}{2}\Delta_{\rm b}(\vx) |\tilde{\vu}_{\rm turb}(w,\vx)|^2.
\end{flalign}
Note that here we do not use $\rho_{\rm b}(\vx)$ directly, but instead use $\Delta_{\rm b}(\vx) = \rho_{\rm b} (\vx)/\bar{\rho}_{\rm b}$ for the calculations, where $\bar{\rho}_{\rm b}$ is the background baryonic density.

\begin{figure*}
\centerline{\includegraphics[width=0.975\textwidth]{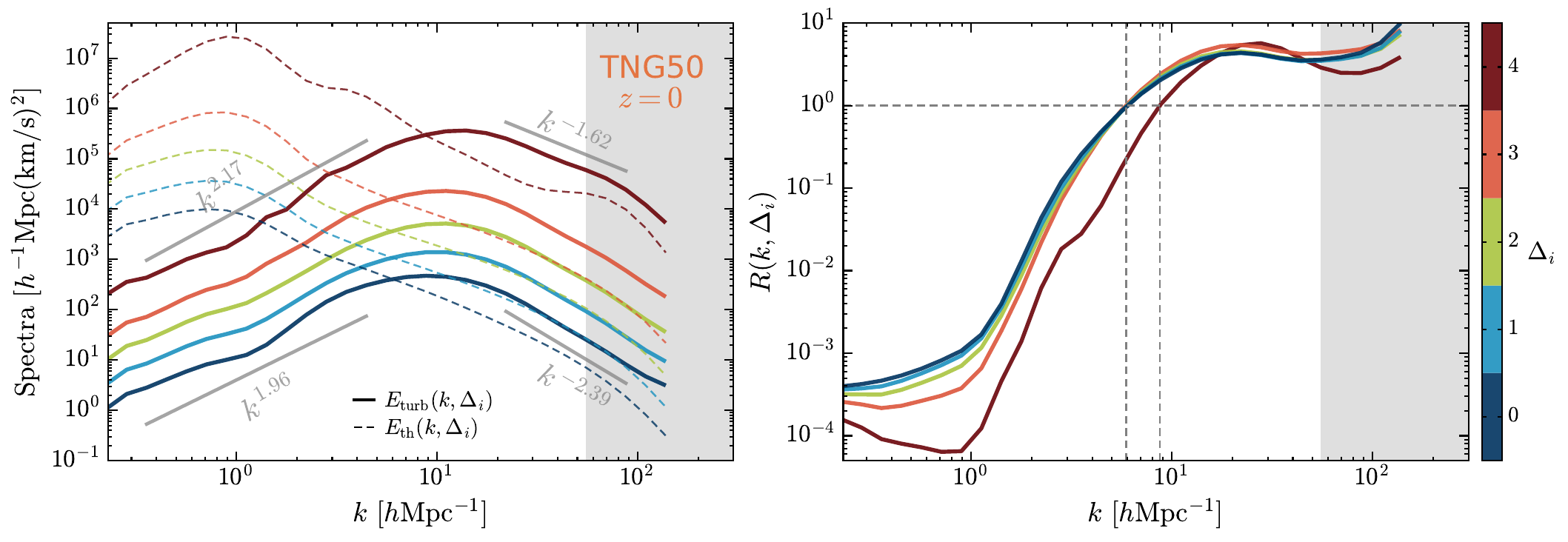}}
\caption{The env-dependent energy spectrum (left panel) and the spectral ratio (right panel) for the turbulent and thermal motions in the TNG50 simulation at $z=0$. The energy spectrum $E(k,\delta)$ is obtained from the env-WPS in Equation~(\ref{eq:env-wps}) as $E(k, \delta) = k^2 S(k, \delta)$. As in Figure~\ref{fig:TNG50_ES_global_evolution}, the relevant power-law fits for the scale ranges that are smaller and larger than the peak positions are indicated in the figure. The vertical dashed lines indicate $k_{\rm equal}$, where $R(k_{\rm equal}) = 1$. The grey area shows the $k > 0.4 k_{\rm Nyquist}$ region.}
\label{fig:TNG50_ES_env_z0}
\end{figure*}

For the the internal energy per unit mass of baryonic matter $e_{\rm u}(\vx)$, we can define a `thermal velocity' as
\begin{flalign}
\label{eq:thermo-velocity}
u_{\rm th}(\vx) \equiv \sqrt{e_{\rm u}(\vx)}.
\end{flalign}
Based on Equation~(\ref{eq:thermo-energy}) and with reference to Equation~(\ref{eq:spect-turb}), we define the local WPS for the thermal velocity $u_{\rm th}$ as
\begin{flalign}
\label{eq:spect-thermo}
S_{\rm th}(w, \vx) \equiv \Delta_{\rm b}(\vx) |\tilde{u}_{\rm th}(w,\vx)|^2,
\end{flalign}
where $\tilde{u}_{\rm th}(w,\vx)$ is the wavelet transform of $u_{\rm th}(\vx)$. Then we define the global WPS for both turbulent and thermal motion as
\begin{flalign}
\label{eq:global-wps}
S_{\rm turb}(w) & \equiv \frac{1}{2}\left<\Delta_{\rm b}(\vx) |\tilde{\vu}_{\rm turb}(w,\vx)|^2\right>_\vx, \nonumber \\
S_{\rm th}(w) & \equiv \left<\Delta_{\rm b}(\vx) |\tilde{u}_{\rm th}(w,\vx)|^2\right>_\vx,
\end{flalign}
in which the average $\left<...\right>$ is performed over all spatial coordinates $\vx$, and the env-WPS as
\begin{flalign}
\label{eq:env-wps}
S_{\rm turb}(w, \delta) & \equiv \frac{1}{2}\left<\Delta_{\rm b}(\vx) |\tilde{\vu}_{\rm turb}(w, \vx)|^2\right>_{\delta(\vx)=\delta}, \nonumber \\
S_{\rm th}(w, \delta) & \equiv \left<\Delta_{\rm b}(\vx) |\tilde{u}_{\rm th}(w,\vx)|^2\right>_{\delta(\vx)=\delta},
\end{flalign}
where the environment is specified with the density contrast $\delta$ and the average $\left<...\right>_{\delta(\vx)=\delta}$ is performed over all the spatial points where the condition ${\delta(\vx)=\delta}$ is satisfied. Note that in \citet{Wang2024b}, we only define the WPS of the velocity field $\vu$. Here, with the prefactor $\Delta_{\rm b}(\vx)$, in Equations~(\ref{eq:global-wps}) and (\ref{eq:env-wps}) we actually define the density-weighted WPS for turbulent and thermal motions. With reference to Equations~(\ref{eq:thermo-energy}) and (\ref{eq:turb-energy}), Equations~(\ref{eq:global-wps}) and (\ref{eq:env-wps}) can also be regarded as spectra for turbulent and thermal energy, respectively. Furthermore, as suggested by Equations~(\ref{eq:thermo-pressure}) and (\ref{eq:turb-pressure}), if they are multiplied by a prefactor of $2/3$, they also represent the spectra for turbulent and thermal pressure, respectively.

We refer readers to Section 2.1 of \citet{Wang2024b} for details on the computation of WPS. Furthermore, in addition to our method, the power spectrum of kinetic or thermal energy can also be computed using the definition of the density-weighted velocity, ${\bf w} \equiv \sqrt{\rho_{\rm b}} \vu$ \citep[e.g.,][]{Kritsuk2007, Schmidt2019}. A comparison of these two methods for computing the power spectrum yields consistent results (not shown).

With Equation~(\ref{eq:global-wps}), we can define the ratio of the global WPS as
\begin{flalign}
\label{eq:global-ratio}
R(k) \equiv \frac{S_{\rm turb}(k)}{S_{\rm th}(k)} = \frac{\left<\Delta_{\rm b}(\vx) |\tilde{\vu}_{\rm turb}(k, \vx)|^2\right>_\vx}{2\left<\Delta_{\rm b}(\vx) |\tilde{u}_{\rm th}(k, \vx)|^2\right>_\vx},
\end{flalign}
and with Equation~(\ref{eq:env-wps}), the ratio of the env-WPS as
\begin{flalign}
\label{eq:env-ratio}
R(k, \delta) \equiv \frac{S_{\rm turb}(k, \delta)}{S_{\rm th}(k, \delta)} = \frac{\left<\Delta_{\rm b}(\vx) |\tilde{\vu}_{\rm turb}(k, \vx)|^2\right>_{\delta(\vx)=\delta}}{2\left<\Delta_{\rm b}(\vx) |\tilde{u}_{\rm th}(k, \vx)|^2\right>_{\delta(\vx)=\delta}},
\end{flalign}
In Equations~(\ref{eq:global-ratio}) and (\ref{eq:env-ratio}), the correspondence $w = c_{w} k$ is used \citep[see][]{Wang2024a, Wang2024b}. As analysed above, the ratios in Equations~(\ref{eq:global-ratio}) and Equations~(\ref{eq:env-ratio}) can be also considered as the ratio of turbulent pressure to thermal pressure.

For details of the CWT techniques that we have developed, please refer to \citet{Wang2021}, \citet{Wang2022a}, \cite{Wang2022b, Wang2023} and \citep{Wang2024a, Wang2024b}.


As analysed in \citet{Wang2024a, Wang2024b}, the numerical effects such as smearing, aliasing and shot noise, are significant when $k > 0.4 k_{\rm Nyquist}$, where $k_{\rm Nyquist}$ is the Nyquist frequency, and hence we restrict ourselves to consider only the scale range of $k < 0.4 k_{\rm Nyquist}$.

In Figure~\ref{fig:TNG50_ES_global_evolution}, we show the $z$-evolution of the global pressure spectra for both turbulent and thermal motions, as well as the spectral ratios $R(k)$. Similar to the results of \citet{Wang2024b}, it can be seen that there are also peaks in all the density-weighted energy spectra for the four redshifts. We observe that the turbulent spectrum grows slightly from $z=4$ to $2$, but increases significantly from $z=2$ to $1$ in the small $k$ or large-scale range, indicating that turbulence induced by structure formation is increasingly injected into the cosmic fluid with time. However, the spectrum remains almost unchanged from $z=1$ to $0$, indicating that the turbulence is close to saturation, or that the injection of turbulence by structure formation is balanced by the transfer from the larger to the smaller scales and the dissipation of turbulence into heat.

From the spectral ratios, we see that $R(k)$ roughly is an increasing function with increasing $k$ when $k>1\hmpc$, indicating turbulent pressure becomes increasingly larger with decreasing scales. At $k_{\rm equal} = 8.1\hmpc$ (or $\lambda_{\rm equal} = 2\pi/k_{\rm equal}=0.78\mpch$) of $z=1$ and $z=0$, the turbulent pressure is comparable to the thermal pressure, whereas when $k>k_{\rm equal}$, the turbulent pressure is dominant over the thermal pressure. The magnitude of turbulent pressure can be as high as $\sim6$ times that of the thermal pressure. Similar conclusions also apply to the cases of $z=4$ and $z=2$, except that in these instances, $k_{\rm equal} = 15.5\hmpc$ (or $\lambda_{\rm equal} = 0.41\mpch$).

In \citet{Wang2024b}, we predict that all WPSs at small scales of the cosmic velocity field are steeper than not only the Kolmogorov exponent $-5/3$ but also the Burgers turbulence exponent $-2$. Here, we observe that the slopes of the turbulent pressure spectrum are $-1.74$ for $z=0$ and $-1.99$ for $z=4$, respectively, which are gentler than those reported in the previous work.

\begin{table}
\centering

\caption{The local density environments, specified with $\Delta_\mathrm{dm}=\rho_\mathrm{dm}/\bar\rho_\mathrm{dm}$ of dark matter.}
\begin{tabular}{lccccc}
\hline
    & $\Delta_0$ & $\Delta_1$ & $\Delta_2$ & $\Delta_3$ & $\Delta_4$  \\
\hline
$\Delta_\mathrm{dm}\in$ & $[0, \ 1/8)$ & $[1/8, \ 1/2)$ & $[1/2, \ 2)$  & $[2, \ 8)$ & $[8, \ +\infty)$  \\
\hline
\end{tabular}
\label{tab:dens_envs}
\end{table}

As in \citet{Wang2024b}, we divide the simulation space of TNG50 into five different environments according to the dark matter density, denoted as $\Delta_i$ with $i=0, 1, ..., 4$ (see Table~\ref{tab:dens_envs}). Among them, $\Delta_0$ and $\Delta_1$ can be regarded as voids or low-density regions, and $\Delta_4$ as various high-density structures, such as clusters, filaments, sheets and their outskirts. In Figure~\ref{fig:TNG50_ES_env_z0}, we present the environment-dependent pressure spectra at $z=0$. One can see that the magnitude of the turbulent pressure spectrum increases with increasing environmental density, but the slope of the spectrum in the small-scale regime becomes increasingly gentle as the density increases. The spectral ratios of turbulent pressure to thermal pressure $R(k)$ are similar for all environmental densities, except that $k_{\rm equal} = 8.7\hmpc$ for the highest density $\Delta_4$, whereas $k_{\rm equal} = 5.9\hmpc$ for other densities. We also find that the maximum spectral ratio is $\sim4$ at $k=20\hmpc$ for the lowest density $\Delta_0$, and as high as $\sim6$ at $k=28\hmpc$ for the highest density $\Delta_4$.

\begin{figure}
\centerline{\includegraphics[width=0.475\textwidth]{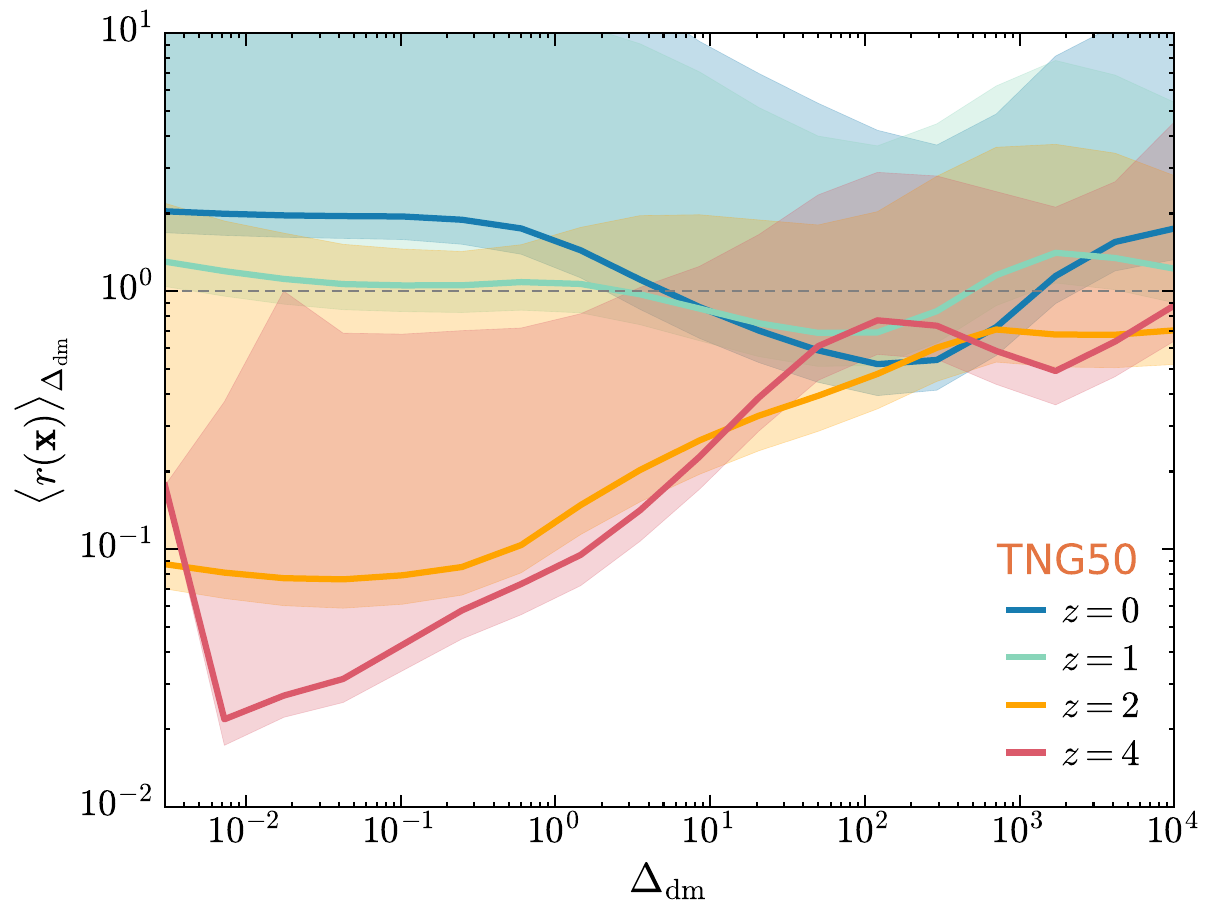}}
\caption{The $z$-evolution of the pressure ratio of turbulent pressure to thermal pressure with respect to dark matter density. The lines represent the mean values of the raio $r(\vx)$, and the color bands indicate the standard deviations for the values above and below the mean calculated separately.}
\label{fig:pressure_ratio}
\end{figure}

In Figure~\ref{fig:pressure_ratio}, we show the $z$-evolution of the pressure ratio of turbulent pressure to thermal pressure with respect to dark matter density. For a spatial location $\vx$, there exists the correspondence between the pressure ratio $r(\vx)$, which is evaluated with Equation~(\ref{eq:pressure-ratio}), and the dark matter density $\Delta_{\rm dm}(\vx) = \rho_{\rm dm}(\vx)/\bar{\rho}_{\rm dm}$. For a density interval centered at $\Delta_{\rm dm}$, we count all the points within this interval, and in this way we can calculate the mean value and the standard deviation of $r(\vx)$. In general, the ratio of turbulent pressure to thermal pressure increases as the redshift decreases. At low redshifts, the ratio in high-density regions is greater than one, and can even reach two times as high as that in lower-density regions. Since these high-density regions are essentially the centers of clusters and filaments, it suggests that turbulent pressure dominates over thermal pressure in these central regions. This dominance is already illustrated by the spectral ratios in Figures~\ref{fig:TNG50_ES_global_evolution} and \ref{fig:TNG50_ES_env_z0}, and this result is also in agreement with the findings of \citet{Vazza2011}, \citet{Parrish2012}, \citet{Iapichino2013} and \citet{Schmidt2017}.

\begin{figure*}
\centerline{
  \includegraphics[width=0.475\textwidth]{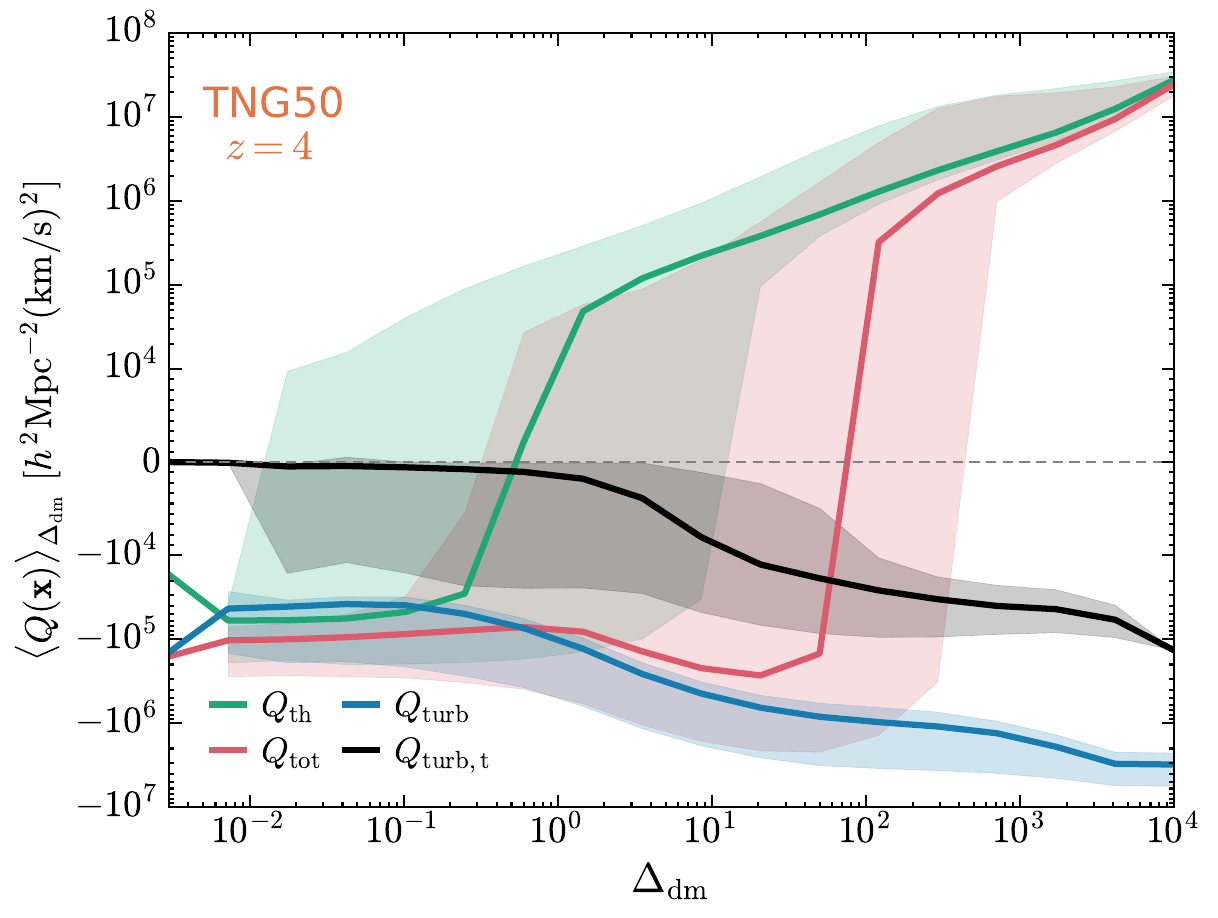}
  \includegraphics[width=0.475\textwidth]{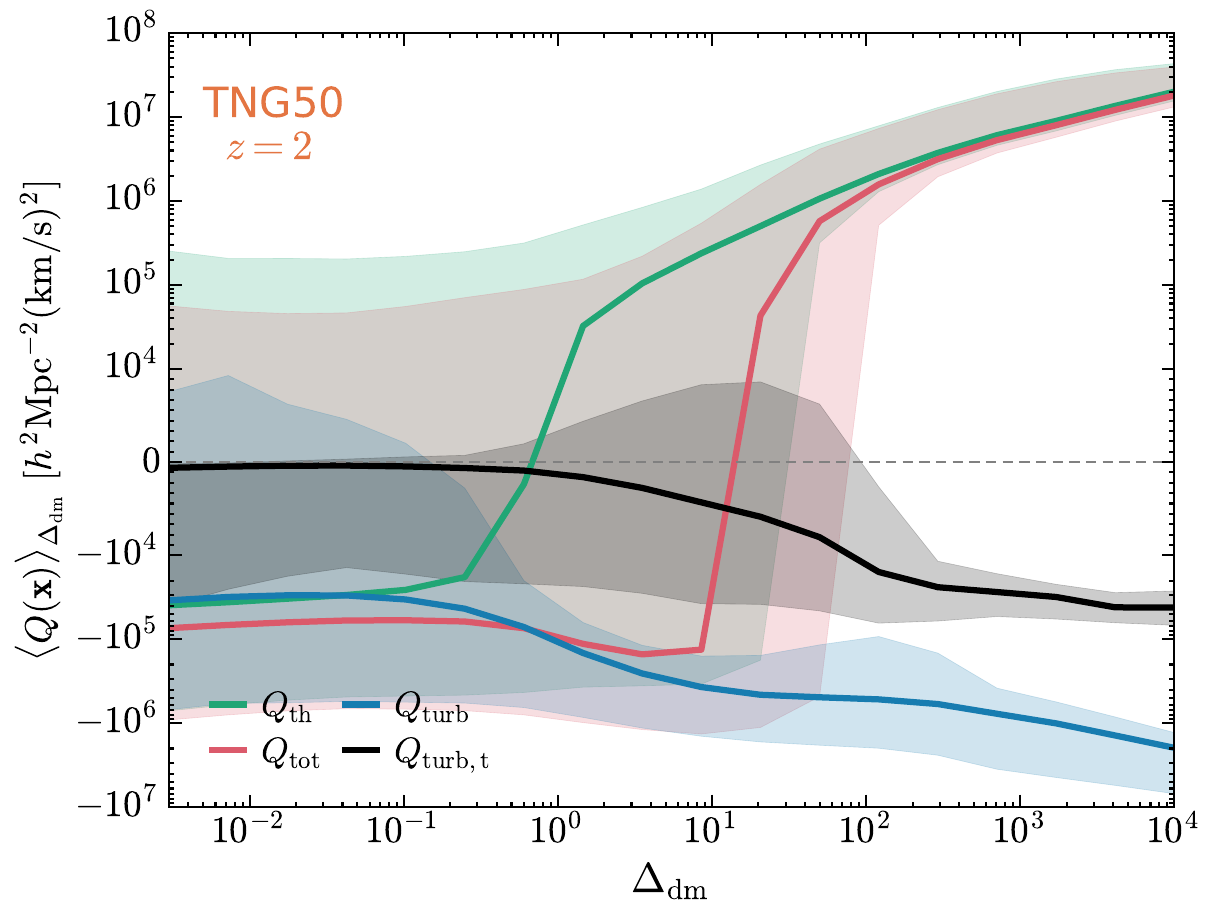}}
\centerline{
  \includegraphics[width=0.475\textwidth]{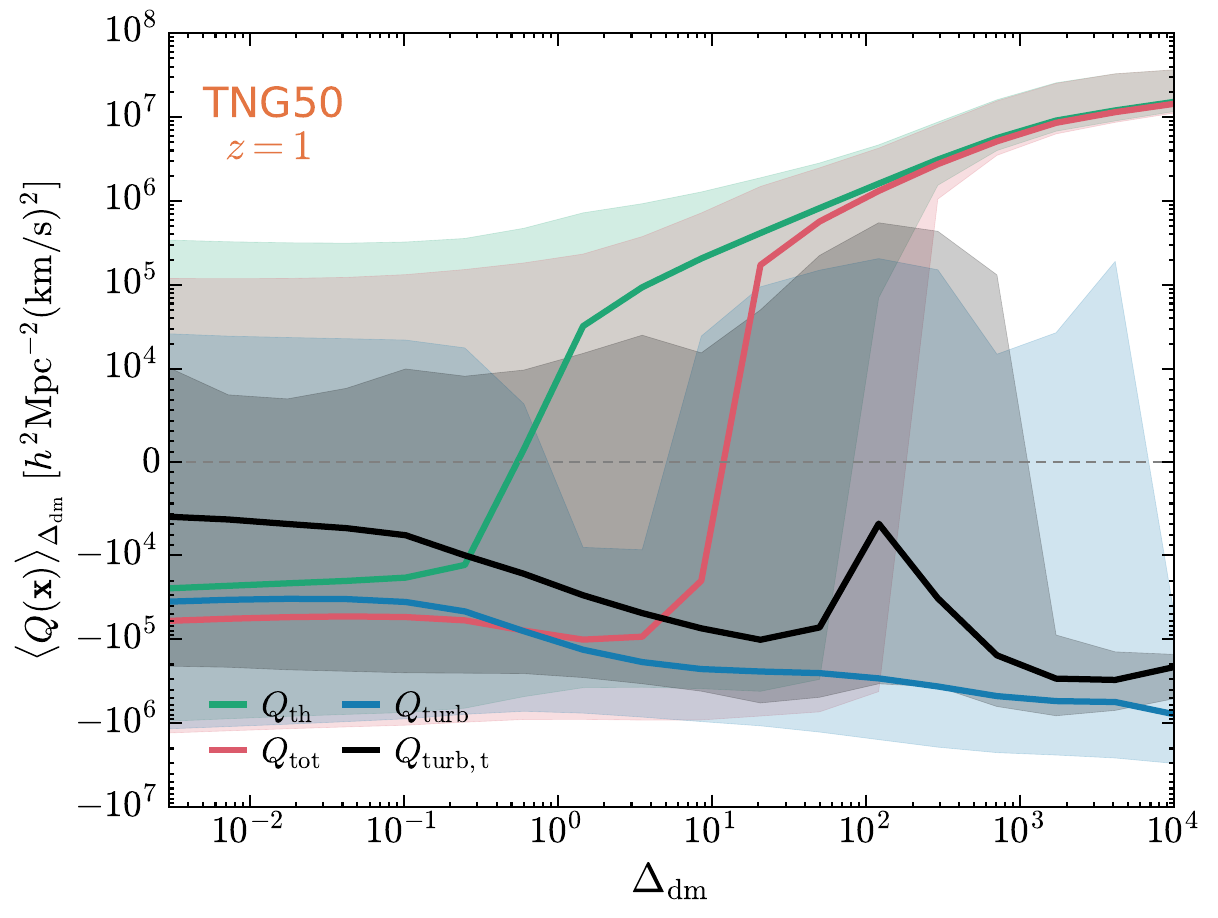}
  \includegraphics[width=0.475\textwidth]{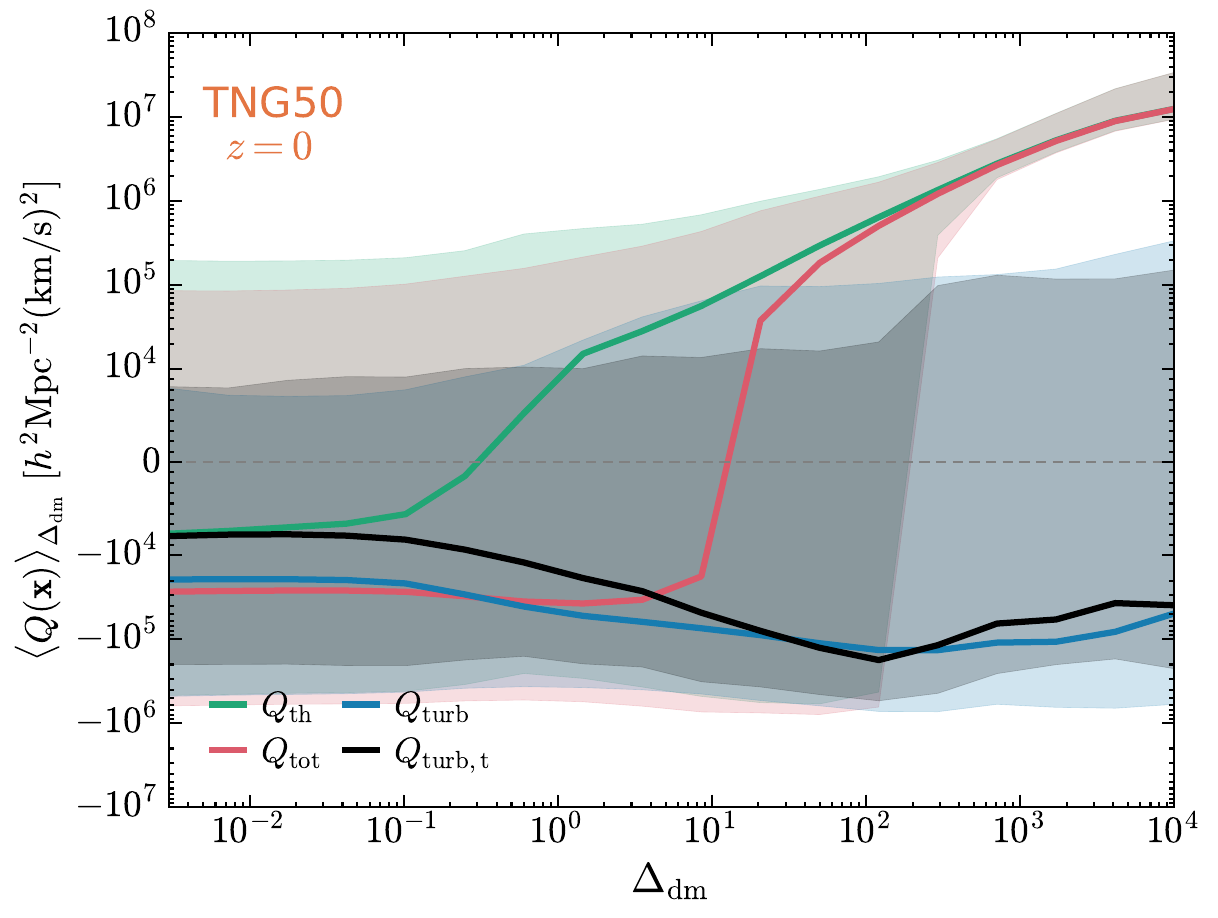}}
\caption{The dynamical effects of turbulence and thermal motion in the cosmic fluid, as well as their combined effects, are represented by $Q_{\rm turb}$, $Q_{\rm turb, t}$, $Q_{\rm th}$ and $Q_{\rm tot}$, respectively. Among these, $Q_{\rm turb}$ and $Q_{\rm turb, t}$ are computed using the total velocity $\vu$ and the bulk flow-removed velocity $\vu_{\rm turb}$, respectively.} The four panels, from top-left to bottom-right, correspond to $z=4$, $2$, $1$, and $0$, respectively. The lines represent the mean values of $Q(\vx)$, and the color bands indicate the standard deviations for the values above and below the mean calculated separately.
\label{fig:Q_vs_dens_evolution}
\end{figure*}

\section{Dynamical Effects}
\label{sec:dynamics}

In this section, we examine in detail the dynamical effects of turbulence and thermal motions in the cosmic baryonic fluid and their combined impact on structure formation. For the velocity field $\vu$ of the cosmic fluid, the dynamical equation of its divergence $d=\nabla\cdot\vu$, or the compression rate, is as follows \citep{Zhu2010, Zhu2011a, Schmidt2017}:
\begin{flalign}
\label{eq:compress-rate}
\frac{Dd}{Dt} \equiv \frac{\partial d}{\partial t} + \frac{1}{a}\vu\cdot\nabla{d} = Q_{\rm turb} + Q_{\rm th} + Q_{\rm grav} + Q_{\rm exp},
\end{flalign}
with
\begin{flalign}
\label{eq:Q-terms}
Q_{\rm turb} & = \frac{1}{a}\left(\frac{1}{2}\omega^2-S^2_{ij}\right), \nonumber \\
Q_{\rm th} & =  \frac{1}{a}\left(\frac{1}{\rho^2_{\rm b}}\nabla\rho_{\rm b}\cdot\nabla{P_{\rm th}} - \frac{1}{\rho_{\rm b}}\nabla^2 {P_{\rm th}} \right), \nonumber \\
Q_{\rm grav} & = -\frac{4\pi G}{a^2}\left( \rho_{\rm tot} -\bar{\rho}_0 \right), \nonumber \\
Q_{\rm exp} & = -\frac{\dot{a}}{a}d,
\end{flalign}
in which $Q_{\rm turb}$, $Q_{\rm th}$, $Q_{\rm grav}$ and $Q_{\rm exp}$ indicate the dynamical effects of turbulence, thermal, gravitational collapse and cosmic expansion, respectively. In the above expression, $\boldsymbol{\omega} \equiv \nabla\times\vu$ is the vorticity, and $S_{ij} \equiv (1/2)({\partial_i}{u_j} + {\partial_j}{u_i})$ is the strain rate of the cosmic fluid. The physical effects of these $Q$ terms are clear, i.e. when $Q>0$, the corresponding effect drives the spatial region or the cosmic structure to expand, whereas $Q<0$, it drives the region/structure to collapse. Here, we focus only on the combined effects of turbulent and thermal dynamics, with
\begin{flalign}
\label{eq:Q-tot}
Q_{\rm tot} = Q_{\rm turb} + Q_{\rm th}.
\end{flalign}
Notice that $Q_{\rm turb}$ should be computed using the total velocity $\vu$, but in order to clearly understand the role of turbulence and its contribution to the compression rate, we also use the bulk flow-removed $\vu_{\rm turb}$ to calculate this dynamic effect, with the result denoted as $Q_{\rm turb, t}$.

In Figure~\ref{fig:Q_vs_dens_evolution}, we present the dynamical effects of turbulence $Q_{\rm turb}$ and $Q_{\rm turb, t}$, thermal motion $Q_{\rm th}$ and their combine effects $Q_{\rm tot}$ as a function of dark matter density $\Delta_{\rm dm}$. For every spatial position $\vx$, there is a correspondence between $Q(\vx)$ and the dark matter density $\Delta_{\rm dm}(\vx)$. As done for Figure~\ref{fig:pressure_ratio}, for a density interval centered at $\Delta_{\rm dm}$, we count all the points within this interval, and in this way we can calculate the mean value and the standard deviation of $Q(\vx)$. We observe that for all four redshifts, in the low-density regions, $Q_{\rm turb}$, $Q_{\rm th}$, and $Q_{\rm tot}$ are all less than zero, exhibiting a considerable dispersion. Moreover, at low redshifts, $Q_{\rm tot}$ is predominantly influenced by $Q_{\rm turb}$. These negative dynamical effects suggest that both turbulent and thermal pressure effects are likely to induce the collapse of the low-density regions. Moreover, these trends towards collapse do not change with redshift. However, when we focus on the high-density regions, we find that the dynamical effects of turbulent pressure and thermal pressure are exactly opposite, in that $Q_{\rm th} > 0$ but $Q_{\rm turb} < 0$, and their combined effect $Q_{\rm tot} > 0$, which is consistent with the dynamical effect of thermal pressure. This conclusion is completely valid for high redshifts ($z=4$ and $2$), but for low redshifts ($z=1$ and $0$), because of the large dispersion for $Q_{\rm turb}$, there is also a small possibility that $Q_{\rm turb}>0$ in some spatial regions. 

As discussed earlier, $Q_{\rm th} > 0$ implies that the thermal pressure will cause the region to expand, while $Q_{\rm turb}<0$ indicates that the turbulent pressure will cause the region to collapse. It is widely accepted that turbulent pressure contributes to the thermal support of the gas, thereby enhancing the effective pressure support and preventing the overcooling problem. Here, we show that this viewpoint is incorrect. In fact, we notice that when $\Delta_{\rm dm}>100$, $Q_{\rm tot}\approx Q_{\rm th} > 0$, indicating that the pressure support within clusters is primarily provided by thermal pressure alone; while turbulent pressure, in almost all cases, effectively counteracts this support.

As mentioned previously, $Q_{\rm turb}$ represents the combined effects of both bulk flow and turbulence. Therefore, it does not directly reflect the contribution of turbulence alone. We may be concerned about the role played by turbulence after the bulk flow is removed. This can be examined by analyzing $Q_{\rm turb, t}$ as a function of dark matter density $\Delta_{\rm dm}$ and its evolution with redshift. From Figure~\ref{fig:Q_vs_dens_evolution}, we see that for all redshifts, $Q_{\rm turb, t}$ contributes negatively to the compression rate. At high redshifts, bulk flow dominates turbulent flow, whereas as redshift decreases, turbulence becomes increasingly important. At $z=0$, turbulence makes the dominant contribution, resulting in $Q_{\rm turb, t} \approx Q_{\rm turb}$ in high-density regions. These results completely confirm the above discussions about $Q_{\rm turb}$, indicating that turbulent pressure effectively counteracts the support provided by thermal pressure.

\section{Conclusions and Discussions}
\label{sec:concl}

Following \citet{Wang2024b}, we apply the CWT techniques to the TNG data to investigate the turbulent and thermal motions of the cosmic baryonic fluid in the large-scale structures of the Universe. Due to the abundance of the included physical processes and AREPO's excellent performance in hydrodynamical computations, the TNG data are well-suited for studying cosmic fluid turbulence.

By defining the pressure spectra, or density-weighted velocity power spectra, as well as the spectral ratios, for both turbulent and thermal motions, we investigate the $z$-evolution of these spectra and the dependence of the spectral ratios on wavenumber $k$. We also study the contributions of turbulent and thermal pressure to the compression rate by defining their respective dynamical effects as $Q_{\rm turb}$ and $Q_{\rm th}$.

Our main findings and conclusions are as follows:
\begin{enumerate}
  \item Similar to the turbulent velocity spectra presented by \citet{Wang2024b}, peaks also exist in the turbulent pressure spectra for all redshifts. The turbulent spectrum shows slight growth from $z=4$ to $2$ and a significant increase from $z=2$ to $1$ at large scales, suggesting that turbulence is progressively introduced into the cosmic fluid. From $z=1$ to $0$, the spectrum remains nearly constant, indicating that turbulence may be nearing saturation or is balanced by energy transfer and dissipation.
  \item The spectral ratio $R(k)$ increases with $k$ for $k > 1\hmpc$, indicating that turbulent pressure becomes more significant at smaller scales. At a specific $k_{\rm equal}$, the turbulent pressure is comparable to thermal pressure, but dominates at larger $k$. The turbulent pressure can exceed the thermal pressure by up to approximately 6 times.
  \item The magnitude of the turbulent pressure spectrum increases with environmental density. The spectral ratios $R(k,\Delta)$ are similar across different environmental densities, with a maximum ratio of approximately 4 to 6 depending on the density. The $z$-evolution of pressure spectra and spectral ratios in different environments generally follows the trend of global spectra. 
  \item The environment-dependent pressure spectra in low-density regions have steeper slopes at small scales compared to high-density regions, indicating more efficient energy transfer in low-density environments\footnote{ See Appendix-B of \citet{Wang2024b}, where we provide a detailed explanation about the relationship between exponent of energy spectrum and energy transfer rate for turbulence.}. Due to the density-weighted prefactor $\Delta_{\rm b}$, the slopes of the turbulent pressure spectrum at large $k$ are less steep than that of the velocity power spectrum of \citet{Wang2024b}.
  \item Generally, the ratio of turbulent pressure to thermal pressure increases as the redshift decreases. At low redshifts, the ratio in high-density regions is greater than one, and can even reach two times as high as that in lower-density regions. Since these high-density regions are essentially the centers of clusters and filaments, it indicates that the turbulent pressure is dominant over the thermal pressure in these central regions, which is consistent with that of \citet{Vazza2011}, \citet{Parrish2012}, \citet{Iapichino2013} and \citet{Schmidt2017}.
  \item For the four redshifts, low-density regions exhibit negative values for turbulence ($Q_{\rm turb}$), thermal motion ($Q_{\rm th}$), and their combined effects ($Q_{\rm tot}$), suggesting a collapse tendency due to both turbulent and thermal pressures, with no significant redshift-related changes. In contrast, high-density regions exhibit the opposite effects: $Q_{\rm th} > 0$ promotes expansion, while $Q_{\rm turb}<0$ leads to collapse, with the combined effect $Q_{\rm tot} > 0$ aligning with the thermal pressure. This conclusion is completely valid for high redshifts, but for low redshifts, because of the large dispersion for $Q_{\rm turb}$, there is also a small possibility that $Q_{\rm turb}>0$ in some spatial regions. Analysis based on $Q_{\rm turb, t}$, which is computed using turbulent velocity, indicates that at $z=0$, turbulence indeed makes the dominant contribution to the compression rate, resulting in $Q_{\rm turb, t} \approx Q_{\rm turb}$ in high-density regions, which completely confirms the previous discussions about $Q_{\rm turb}$.
\end{enumerate}

The Kolmogorov turbulence, which is homogeneous and isotropic in space and is characterized by eddies of different scales, is the simple theoretical framework for turbulence. However, this picture is problematic for the turbulence of the cosmic baryonic fluid, which is subject to structure formation driven by gravity in the context of cosmic expansion. Structure formation leads, for example, to the density stratification of the cosmic fluid, where buoyancy forces resist radial motions, making the turbulence inhomogeneous and anisotropic \citep[e.g.][]{Shi2018, Shi2019, Mohapatra2020, Simonte2022, Wangc2023}. In this study, the density-weighted prefactor $\Delta_{\rm b}$ in the definition of the pressure power spectrum inherently accounts for the inhomogeneity and anisotropy of turbulent motions in the cosmic fluid, and hence the issues of density stratification are partially resolved.

Contrary to the common belief that turbulent pressure provides extra pressure support to prevent overcooling, the data indicate that when $\Delta_{\rm dm} > 100$, $Q_{\rm turb}<0$ and $Q_{\rm tot} \approx Q_{\rm th}>0$, revealing that thermal pressure alone is the main source of pressure support within clusters. Turbulent pressure, in most cases, works against this support, challenging the previously held viewpoint. This conclusion aligns with \citet{Schmidt2017}'s, but their conclusion is based on a specific cluster selected from their simulation sample. In contrast, our results, derived from a statistical analysis of the TNG data, are more general and universal.

\section*{Acknowledgments}

The authors thank the anonymous referee for helpful comments and suggestions. We acknowledge the use of the data from IllustrisTNG simulation for this work. We also acknowledge the support by the National Science Foundation of China (No. 12147217, 12347163), and by the Natural Science Foundation of Jilin Province, China (No. 20180101228JC).

\section*{Data Availability}
The data used in this paper are available from the correspondence author upon reasonable request.

\bibliography{references}{}

\begin{thebibliography}{}
\makeatletter
\relax
\def\mn@urlcharsother{\let\do\@makeother \do\$\do\&\do\#\do\^\do\_\do\%\do\~}
\def\mn@doi{\begingroup\mn@urlcharsother \@ifnextchar [ {\mn@doi@} {\mn@doi@[]}}
\def\mn@doi@[#1]#2{\def\@tempa{#1}\ifx\@tempa\@empty \href {http://dx.doi.org/#2} {doi:#2}\else \href {http://dx.doi.org/#2} {#1}\fi \endgroup}
\def\mn@eprint#1#2{\mn@eprint@#1:#2::\@nil}
\def\mn@eprint@arXiv#1{\href {http://arxiv.org/abs/#1} {{\tt arXiv:#1}}}
\def\mn@eprint@dblp#1{\href {http://dblp.uni-trier.de/rec/bibtex/#1.xml} {dblp:#1}}
\def\mn@eprint@#1:#2:#3:#4\@nil{\def\@tempa {#1}\def\@tempb {#2}\def\@tempc {#3}\ifx \@tempc \@empty \let \@tempc \@tempb \let \@tempb \@tempa \fi \ifx \@tempb \@empty \def\@tempb {arXiv}\fi \@ifundefined {mn@eprint@\@tempb}{\@tempb:\@tempc}{\expandafter \expandafter \csname mn@eprint@\@tempb\endcsname \expandafter{\@tempc}}}

\bibitem[\protect\citeauthoryear{{Angelinelli}, {Vazza}, {Giocoli}, {Ettori}, {Jones}, {Brunetti}, {Br{\"u}ggen}  \& {Eckert}}{{Angelinelli} et~al.}{2020}]{Angelinelli2020}
{Angelinelli} M.,  {Vazza} F.,  {Giocoli} C.,  {Ettori} S.,  {Jones} T.~W.,  {Brunetti} G.,  {Br{\"u}ggen} M.,   {Eckert} D.,  2020, \mn@doi [\mnras] {10.1093/mnras/staa975}, \href {https://ui.adsabs.harvard.edu/abs/2020MNRAS.495..864A} {495, 864}

\bibitem[\protect\citeauthoryear{{Banerjee} \& {Sharma}}{{Banerjee} \& {Sharma}}{2014}]{Banerjee2014}
{Banerjee} N.,  {Sharma} P.,  2014, \mn@doi [\mnras] {10.1093/mnras/stu1179}, \href {https://ui.adsabs.harvard.edu/abs/2014MNRAS.443..687B} {443, 687}

\bibitem[\protect\citeauthoryear{{Bauer} \& {Springel}}{{Bauer} \& {Springel}}{2012}]{Bauer2012}
{Bauer} A.,  {Springel} V.,  2012, \mn@doi [\mnras] {10.1111/j.1365-2966.2012.21058.x}, \href {https://ui.adsabs.harvard.edu/abs/2012MNRAS.423.2558B} {423, 2558}

\bibitem[\protect\citeauthoryear{{Bonafede}, {Feretti}, {Murgia}, {Govoni}, {Giovannini}, {Dallacasa}, {Dolag}  \& {Taylor}}{{Bonafede} et~al.}{2010}]{Bonafede2010}
{Bonafede} A.,  {Feretti} L.,  {Murgia} M.,  {Govoni} F.,  {Giovannini} G.,  {Dallacasa} D.,  {Dolag} K.,   {Taylor} G.~B.,  2010, \mn@doi [\aap] {10.1051/0004-6361/200913696}, \href {https://ui.adsabs.harvard.edu/abs/2010A&A...513A..30B} {513, A30}

\bibitem[\protect\citeauthoryear{{Bonazzola}, {Heyvaerts}, {Falgarone}, {Perault}  \& {Puget}}{{Bonazzola} et~al.}{1987}]{Bonazzola1987}
{Bonazzola} S.,  {Heyvaerts} J.,  {Falgarone} E.,  {Perault} M.,   {Puget} J.~L.,  1987, \aap, \href {https://ui.adsabs.harvard.edu/abs/1987A&A...172..293B} {172, 293}

\bibitem[\protect\citeauthoryear{{Br{\"u}ggen} \& {Vazza}}{{Br{\"u}ggen} \& {Vazza}}{2015}]{Bruggen2015}
{Br{\"u}ggen} M.,  {Vazza} F.,  2015, in {Lazarian} A.,  {de Gouveia Dal Pino} E.~M.,   {Melioli} C.,  eds,  Astrophysics and Space Science Library Vol. 407, Magnetic Fields in Diffuse Media. p.~599, \mn@doi{10.1007/978-3-662-44625-6_21}

\bibitem[\protect\citeauthoryear{{Br{\"u}ggen}, {Hoeft}  \& {Ruszkowski}}{{Br{\"u}ggen} et~al.}{2005}]{Bruggen2005}
{Br{\"u}ggen} M.,  {Hoeft} M.,   {Ruszkowski} M.,  2005, \mn@doi [\apj] {10.1086/430732}, \href {https://ui.adsabs.harvard.edu/abs/2005ApJ...628..153B} {628, 153}

\bibitem[\protect\citeauthoryear{{Cassano} \& {Brunetti}}{{Cassano} \& {Brunetti}}{2005}]{Cassano2005}
{Cassano} R.,  {Brunetti} G.,  2005, \mn@doi [\mnras] {10.1111/j.1365-2966.2005.08747.x}, \href {https://ui.adsabs.harvard.edu/abs/2005MNRAS.357.1313C} {357, 1313}

\bibitem[\protect\citeauthoryear{{Chandrasekhar}}{{Chandrasekhar}}{1951a}]{Chandra1951a}
{Chandrasekhar} S.,  1951a, \mn@doi [Proceedings of the Royal Society of London Series A] {10.1098/rspa.1951.0227}, \href {https://ui.adsabs.harvard.edu/abs/1951RSPSA.210...18C} {210, 18}

\bibitem[\protect\citeauthoryear{{Chandrasekhar}}{{Chandrasekhar}}{1951b}]{Chandra1951b}
{Chandrasekhar} S.,  1951b, \mn@doi [Proceedings of the Royal Society of London Series A] {10.1098/rspa.1951.0228}, \href {https://ui.adsabs.harvard.edu/abs/1951RSPSA.210...26C} {210, 26}

\bibitem[\protect\citeauthoryear{{Churazov}, {Forman}, {Jones}, {Sunyaev}  \& {B{\"o}hringer}}{{Churazov} et~al.}{2004}]{Churazov2004}
{Churazov} E.,  {Forman} W.,  {Jones} C.,  {Sunyaev} R.,   {B{\"o}hringer} H.,  2004, \mn@doi [\mnras] {10.1111/j.1365-2966.2004.07201.x}, \href {https://ui.adsabs.harvard.edu/abs/2004MNRAS.347...29C} {347, 29}

\bibitem[\protect\citeauthoryear{{Churazov} et~al.,}{{Churazov} et~al.}{2012}]{Churazov2012}
{Churazov} E.,  et~al., 2012, \mn@doi [\mnras] {10.1111/j.1365-2966.2011.20372.x}, \href {https://ui.adsabs.harvard.edu/abs/2012MNRAS.421.1123C} {421, 1123}

\bibitem[\protect\citeauthoryear{{Dolag}, {Vazza}, {Brunetti}  \& {Tormen}}{{Dolag} et~al.}{2005}]{Dolag2005}
{Dolag} K.,  {Vazza} F.,  {Brunetti} G.,   {Tormen} G.,  2005, \mn@doi [\mnras] {10.1111/j.1365-2966.2005.09630.x}, \href {https://ui.adsabs.harvard.edu/abs/2005MNRAS.364..753D} {364, 753}

\bibitem[\protect\citeauthoryear{{Eckert} et~al.,}{{Eckert} et~al.}{2019}]{Eckert2019}
{Eckert} D.,  et~al., 2019, \mn@doi [\aap] {10.1051/0004-6361/201833324}, \href {https://ui.adsabs.harvard.edu/abs/2019A&A...621A..40E} {621, A40}

\bibitem[\protect\citeauthoryear{{En{\ss}lin} \& {Vogt}}{{En{\ss}lin} \& {Vogt}}{2006}]{Ensslin2006}
{En{\ss}lin} T.~A.,  {Vogt} C.,  2006, \mn@doi [\aap] {10.1051/0004-6361:20053518}, \href {https://ui.adsabs.harvard.edu/abs/2006A&A...453..447E} {453, 447}

\bibitem[\protect\citeauthoryear{{Ettori} et~al.,}{{Ettori} et~al.}{2019}]{Ettori2019}
{Ettori} S.,  et~al., 2019, \mn@doi [\aap] {10.1051/0004-6361/201833323}, \href {https://ui.adsabs.harvard.edu/abs/2019A&A...621A..39E} {621, A39}

\bibitem[\protect\citeauthoryear{{Evoli} \& {Ferrara}}{{Evoli} \& {Ferrara}}{2011}]{Evoli2011}
{Evoli} C.,  {Ferrara} A.,  2011, \mn@doi [\mnras] {10.1111/j.1365-2966.2011.18343.x}, \href {https://ui.adsabs.harvard.edu/abs/2011MNRAS.413.2721E} {413, 2721}

\bibitem[\protect\citeauthoryear{{Fang}, {Humphrey}  \& {Buote}}{{Fang} et~al.}{2009}]{Fangtt2009}
{Fang} T.,  {Humphrey} P.,   {Buote} D.,  2009, \mn@doi [\apj] {10.1088/0004-637X/691/2/1648}, \href {https://ui.adsabs.harvard.edu/abs/2009ApJ...691.1648F} {691, 1648}

\bibitem[\protect\citeauthoryear{{Fusco-Femiano}}{{Fusco-Femiano}}{2019}]{Fusco-Femiano2019}
{Fusco-Femiano} R.,  2019, \mn@doi [\mnras] {10.1093/mnras/stz482}, \href {https://ui.adsabs.harvard.edu/abs/2019MNRAS.485.1800F} {485, 1800}

\bibitem[\protect\citeauthoryear{{Fusco-Femiano} \& {Lapi}}{{Fusco-Femiano} \& {Lapi}}{2014}]{Fusco-Femiano2014}
{Fusco-Femiano} R.,  {Lapi} A.,  2014, \mn@doi [\apj] {10.1088/0004-637X/783/2/76}, \href {https://ui.adsabs.harvard.edu/abs/2014ApJ...783...76F} {783, 76}

\bibitem[\protect\citeauthoryear{{Fusco-Femiano} \& {Lapi}}{{Fusco-Femiano} \& {Lapi}}{2018}]{Fusco-Femiano2018}
{Fusco-Femiano} R.,  {Lapi} A.,  2018, \mn@doi [\mnras] {10.1093/mnras/stx3243}, \href {https://ui.adsabs.harvard.edu/abs/2018MNRAS.475.1340F} {475, 1340}

\bibitem[\protect\citeauthoryear{{Gaspari} \& {Churazov}}{{Gaspari} \& {Churazov}}{2013}]{Gaspari2013}
{Gaspari} M.,  {Churazov} E.,  2013, \mn@doi [\aap] {10.1051/0004-6361/201322295}, \href {https://ui.adsabs.harvard.edu/abs/2013A&A...559A..78G} {559, A78}

\bibitem[\protect\citeauthoryear{{Gaspari}, {Melioli}, {Brighenti}  \& {D'Ercole}}{{Gaspari} et~al.}{2011}]{Gaspari2011}
{Gaspari} M.,  {Melioli} C.,  {Brighenti} F.,   {D'Ercole} A.,  2011, \mn@doi [\mnras] {10.1111/j.1365-2966.2010.17688.x}, \href {https://ui.adsabs.harvard.edu/abs/2011MNRAS.411..349G} {411, 349}

\bibitem[\protect\citeauthoryear{{Gaspari}, {Churazov}, {Nagai}, {Lau}  \& {Zhuravleva}}{{Gaspari} et~al.}{2014}]{Gaspari2014}
{Gaspari} M.,  {Churazov} E.,  {Nagai} D.,  {Lau} E.~T.,   {Zhuravleva} I.,  2014, \mn@doi [\aap] {10.1051/0004-6361/201424043}, \href {https://ui.adsabs.harvard.edu/abs/2014A&A...569A..67G} {569, A67}

\bibitem[\protect\citeauthoryear{{He}, {Liu}, {Feng}, {Shu}  \& {Fang}}{{He} et~al.}{2006}]{Hep2006}
{He} P.,  {Liu} J.,  {Feng} L.-L.,  {Shu} C.-W.,   {Fang} L.-Z.,  2006, \mn@doi [\prl] {10.1103/PhysRevLett.96.051302}, \href {https://ui.adsabs.harvard.edu/abs/2006PhRvL..96e1302H} {96, 051302}

\bibitem[\protect\citeauthoryear{{Hitomi Collaboration} et~al.,}{{Hitomi Collaboration} et~al.}{2016}]{Hitomi2016}
{Hitomi Collaboration} et~al., 2016, \mn@doi [\nat] {10.1038/nature18627}, \href {https://ui.adsabs.harvard.edu/abs/2016Natur.535..117H} {535, 117}

\bibitem[\protect\citeauthoryear{{Hitomi Collaboration} et~al.,}{{Hitomi Collaboration} et~al.}{2018}]{Hitomi2018}
{Hitomi Collaboration} et~al., 2018, \mn@doi [\pasj] {10.1093/pasj/psx127}, \href {https://ui.adsabs.harvard.edu/abs/2018PASJ...70...10H} {70, 10}

\bibitem[\protect\citeauthoryear{{Iapichino} \& {Niemeyer}}{{Iapichino} \& {Niemeyer}}{2008}]{Iapichino2008b}
{Iapichino} L.,  {Niemeyer} J.~C.,  2008, \mn@doi [\mnras] {10.1111/j.1365-2966.2008.13518.x}, \href {https://ui.adsabs.harvard.edu/abs/2008MNRAS.388.1089I} {388, 1089}

\bibitem[\protect\citeauthoryear{{Iapichino}, {Schmidt}, {Niemeyer}  \& {Merklein}}{{Iapichino} et~al.}{2011}]{Iapichino2011}
{Iapichino} L.,  {Schmidt} W.,  {Niemeyer} J.~C.,   {Merklein} J.,  2011, \mn@doi [\mnras] {10.1111/j.1365-2966.2011.18550.x}, \href {https://ui.adsabs.harvard.edu/abs/2011MNRAS.414.2297I} {414, 2297}

\bibitem[\protect\citeauthoryear{{Iapichino}, {Viel}  \& {Borgani}}{{Iapichino} et~al.}{2013}]{Iapichino2013}
{Iapichino} L.,  {Viel} M.,   {Borgani} S.,  2013, \mn@doi [\mnras] {10.1093/mnras/stt611}, \href {https://ui.adsabs.harvard.edu/abs/2013MNRAS.432.2529I} {432, 2529}

\bibitem[\protect\citeauthoryear{{Iapichino}, {Federrath}  \& {Klessen}}{{Iapichino} et~al.}{2017}]{Iapichino2017}
{Iapichino} L.,  {Federrath} C.,   {Klessen} R.~S.,  2017, \mn@doi [\mnras] {10.1093/mnras/stx882}, \href {https://ui.adsabs.harvard.edu/abs/2017MNRAS.469.3641I} {469, 3641}

\bibitem[\protect\citeauthoryear{{Khatri} \& {Gaspari}}{{Khatri} \& {Gaspari}}{2016}]{Khatri2016}
{Khatri} R.,  {Gaspari} M.,  2016, \mn@doi [\mnras] {10.1093/mnras/stw2027}, \href {https://ui.adsabs.harvard.edu/abs/2016MNRAS.463..655K} {463, 655}

\bibitem[\protect\citeauthoryear{{Kritsuk}, {Norman}, {Padoan}  \& {Wagner}}{{Kritsuk} et~al.}{2007}]{Kritsuk2007}
{Kritsuk} A.~G.,  {Norman} M.~L.,  {Padoan} P.,   {Wagner} R.,  2007, \mn@doi [\apj] {10.1086/519443}, \href {https://ui.adsabs.harvard.edu/abs/2007ApJ...665..416K} {665, 416}

\bibitem[\protect\citeauthoryear{{Lau}, {Kravtsov}  \& {Nagai}}{{Lau} et~al.}{2009}]{Lau2009}
{Lau} E.~T.,  {Kravtsov} A.~V.,   {Nagai} D.,  2009, \mn@doi [\apj] {10.1088/0004-637X/705/2/1129}, \href {https://ui.adsabs.harvard.edu/abs/2009ApJ...705.1129L} {705, 1129}

\bibitem[\protect\citeauthoryear{{Marinacci} et~al.,}{{Marinacci} et~al.}{2018}]{Marinacci2018}
{Marinacci} F.,  et~al., 2018, \mn@doi [\mnras] {10.1093/mnras/sty2206}, \href {https://ui.adsabs.harvard.edu/abs/2018MNRAS.480.5113M} {480, 5113}

\bibitem[\protect\citeauthoryear{{Mohapatra}, {Federrath}  \& {Sharma}}{{Mohapatra} et~al.}{2020}]{Mohapatra2020}
{Mohapatra} R.,  {Federrath} C.,   {Sharma} P.,  2020, \mn@doi [\mnras] {10.1093/mnras/staa711}, \href {https://ui.adsabs.harvard.edu/abs/2020MNRAS.493.5838M} {493, 5838}

\bibitem[\protect\citeauthoryear{{Morandi}, {Limousin}, {Rephaeli}, {Umetsu}, {Barkana}, {Broadhurst}  \& {Dahle}}{{Morandi} et~al.}{2011}]{Morandi2011}
{Morandi} A.,  {Limousin} M.,  {Rephaeli} Y.,  {Umetsu} K.,  {Barkana} R.,  {Broadhurst} T.,   {Dahle} H.,  2011, \mn@doi [\mnras] {10.1111/j.1365-2966.2011.19175.x}, \href {https://ui.adsabs.harvard.edu/abs/2011MNRAS.416.2567M} {416, 2567}

\bibitem[\protect\citeauthoryear{{Murgia}, {Govoni}, {Feretti}, {Giovannini}, {Dallacasa}, {Fanti}, {Taylor}  \& {Dolag}}{{Murgia} et~al.}{2004}]{Murgia2004}
{Murgia} M.,  {Govoni} F.,  {Feretti} L.,  {Giovannini} G.,  {Dallacasa} D.,  {Fanti} R.,  {Taylor} G.~B.,   {Dolag} K.,  2004, \mn@doi [\aap] {10.1051/0004-6361:20040191}, \href {https://ui.adsabs.harvard.edu/abs/2004A&A...424..429M} {424, 429}

\bibitem[\protect\citeauthoryear{{Naiman} et~al.,}{{Naiman} et~al.}{2018}]{Naiman2018}
{Naiman} J.~P.,  et~al., 2018, \mn@doi [\mnras] {10.1093/mnras/sty618}, \href {https://ui.adsabs.harvard.edu/abs/2018MNRAS.477.1206N} {477, 1206}

\bibitem[\protect\citeauthoryear{{Nelson} et~al.,}{{Nelson} et~al.}{2018}]{Nelson2018}
{Nelson} D.,  et~al., 2018, \mn@doi [\mnras] {10.1093/mnras/stx3040}, \href {https://ui.adsabs.harvard.edu/abs/2018MNRAS.475..624N} {475, 624}

\bibitem[\protect\citeauthoryear{{Nelson} et~al.,}{{Nelson} et~al.}{2019}]{Nelson2019}
{Nelson} D.,  et~al., 2019, \mn@doi [Computational Astrophysics and Cosmology] {10.1186/s40668-019-0028-x}, \href {https://ui.adsabs.harvard.edu/abs/2019ComAC...6....2N} {6, 2}

\bibitem[\protect\citeauthoryear{{Ota}, {Nagai}  \& {Lau}}{{Ota} et~al.}{2018}]{Ota2018}
{Ota} N.,  {Nagai} D.,   {Lau} E.~T.,  2018, \mn@doi [\pasj] {10.1093/pasj/psy040}, \href {https://ui.adsabs.harvard.edu/abs/2018PASJ...70...51O} {70, 51}

\bibitem[\protect\citeauthoryear{{Parrish}, {McCourt}, {Quataert}  \& {Sharma}}{{Parrish} et~al.}{2012}]{Parrish2012}
{Parrish} I.~J.,  {McCourt} M.,  {Quataert} E.,   {Sharma} P.,  2012, \mn@doi [\mnras] {10.1111/j.1745-3933.2011.01171.x10.5479/ADS/bib/1912LicOB.7.19C}, \href {https://ui.adsabs.harvard.edu/abs/2012MNRAS.419L..29P} {419, L29}

\bibitem[\protect\citeauthoryear{{Paul}, {Iapichino}, {Miniati}, {Bagchi}  \& {Mannheim}}{{Paul} et~al.}{2011}]{Paul2011}
{Paul} S.,  {Iapichino} L.,  {Miniati} F.,  {Bagchi} J.,   {Mannheim} K.,  2011, \mn@doi [\apj] {10.1088/0004-637X/726/1/17}, \href {https://ui.adsabs.harvard.edu/abs/2011ApJ...726...17P} {726, 17}

\bibitem[\protect\citeauthoryear{{Pillepich} et~al.,}{{Pillepich} et~al.}{2018a}]{Pillepich2018a}
{Pillepich} A.,  et~al., 2018a, \mn@doi [\mnras] {10.1093/mnras/stx2656}, \href {https://ui.adsabs.harvard.edu/abs/2018MNRAS.473.4077P} {473, 4077}

\bibitem[\protect\citeauthoryear{{Pillepich} et~al.,}{{Pillepich} et~al.}{2018b}]{Pillepich2018b}
{Pillepich} A.,  et~al., 2018b, \mn@doi [\mnras] {10.1093/mnras/stx3112}, \href {https://ui.adsabs.harvard.edu/abs/2018MNRAS.475..648P} {475, 648}

\bibitem[\protect\citeauthoryear{{Porter}, {Jones}  \& {Ryu}}{{Porter} et~al.}{2015}]{Porter2015}
{Porter} D.~H.,  {Jones} T.~W.,   {Ryu} D.,  2015, \mn@doi [\apj] {10.1088/0004-637X/810/2/93}, \href {https://ui.adsabs.harvard.edu/abs/2015ApJ...810...93P} {810, 93}

\bibitem[\protect\citeauthoryear{{Rasia}, {Tormen}  \& {Moscardini}}{{Rasia} et~al.}{2004}]{Rasia2004}
{Rasia} E.,  {Tormen} G.,   {Moscardini} L.,  2004, \mn@doi [\mnras] {10.1111/j.1365-2966.2004.07775.x}, \href {https://ui.adsabs.harvard.edu/abs/2004MNRAS.351..237R} {351, 237}

\bibitem[\protect\citeauthoryear{{Roediger} \& {Br{\"u}ggen}}{{Roediger} \& {Br{\"u}ggen}}{2007}]{Roediger2007}
{Roediger} E.,  {Br{\"u}ggen} M.,  2007, \mn@doi [\mnras] {10.1111/j.1365-2966.2007.12241.x}, \href {https://ui.adsabs.harvard.edu/abs/2007MNRAS.380.1399R} {380, 1399}

\bibitem[\protect\citeauthoryear{{Ryu}, {Kang}, {Cho}  \& {Das}}{{Ryu} et~al.}{2008}]{Ryu2008}
{Ryu} D.,  {Kang} H.,  {Cho} J.,   {Das} S.,  2008, \mn@doi [Science] {10.1126/science.1154923}, \href {https://ui.adsabs.harvard.edu/abs/2008Sci...320..909R} {320, 909}

\bibitem[\protect\citeauthoryear{{Schmidt} \& {Grete}}{{Schmidt} \& {Grete}}{2019}]{Schmidt2019}
{Schmidt} W.,  {Grete} P.,  2019, \mn@doi [\pre] {10.1103/PhysRevE.100.043116}, \href {https://ui.adsabs.harvard.edu/abs/2019PhRvE.100d3116S} {100, 043116}

\bibitem[\protect\citeauthoryear{{Schmidt}, {Kern}, {Federrath}  \& {Klessen}}{{Schmidt} et~al.}{2010}]{Schmidt2010}
{Schmidt} W.,  {Kern} S.~A.~W.,  {Federrath} C.,   {Klessen} R.~S.,  2010, \mn@doi [\aap] {10.1051/0004-6361/200913904}, \href {https://ui.adsabs.harvard.edu/abs/2010A&A...516A..25S} {516, A25}

\bibitem[\protect\citeauthoryear{{Schmidt}, {Byrohl}, {Engels}, {Behrens}  \& {Niemeyer}}{{Schmidt} et~al.}{2017}]{Schmidt2017}
{Schmidt} W.,  {Byrohl} C.,  {Engels} J.~F.,  {Behrens} C.,   {Niemeyer} J.~C.,  2017, \mn@doi [\mnras] {10.1093/mnras/stx1274}, \href {https://ui.adsabs.harvard.edu/abs/2017MNRAS.470..142S} {470, 142}

\bibitem[\protect\citeauthoryear{{Schuecker}, {Finoguenov}, {Miniati}, {B{\"o}hringer}  \& {Briel}}{{Schuecker} et~al.}{2004}]{Schuecker2004}
{Schuecker} P.,  {Finoguenov} A.,  {Miniati} F.,  {B{\"o}hringer} H.,   {Briel} U.~G.,  2004, \mn@doi [\aap] {10.1051/0004-6361:20041039}, \href {https://ui.adsabs.harvard.edu/abs/2004A&A...426..387S} {426, 387}

\bibitem[\protect\citeauthoryear{{Shi} \& {Komatsu}}{{Shi} \& {Komatsu}}{2014}]{Shi2014}
{Shi} X.,  {Komatsu} E.,  2014, \mn@doi [\mnras] {10.1093/mnras/stu858}, \href {https://ui.adsabs.harvard.edu/abs/2014MNRAS.442..521S} {442, 521}

\bibitem[\protect\citeauthoryear{{Shi} \& {Zhang}}{{Shi} \& {Zhang}}{2019}]{Shi2019}
{Shi} X.,  {Zhang} C.,  2019, \mn@doi [\mnras] {10.1093/mnras/stz1392}, \href {https://ui.adsabs.harvard.edu/abs/2019MNRAS.487.1072S} {487, 1072}

\bibitem[\protect\citeauthoryear{{Shi}, {Komatsu}, {Nelson}  \& {Nagai}}{{Shi} et~al.}{2015}]{Shi2015}
{Shi} X.,  {Komatsu} E.,  {Nelson} K.,   {Nagai} D.,  2015, \mn@doi [\mnras] {10.1093/mnras/stv036}, \href {https://ui.adsabs.harvard.edu/abs/2015MNRAS.448.1020S} {448, 1020}

\bibitem[\protect\citeauthoryear{{Shi}, {Komatsu}, {Nagai}  \& {Lau}}{{Shi} et~al.}{2016}]{Shi2016}
{Shi} X.,  {Komatsu} E.,  {Nagai} D.,   {Lau} E.~T.,  2016, \mn@doi [\mnras] {10.1093/mnras/stv2504}, \href {https://ui.adsabs.harvard.edu/abs/2016MNRAS.455.2936S} {455, 2936}

\bibitem[\protect\citeauthoryear{{Shi}, {Nagai}  \& {Lau}}{{Shi} et~al.}{2018}]{Shi2018}
{Shi} X.,  {Nagai} D.,   {Lau} E.~T.,  2018, \mn@doi [\mnras] {10.1093/mnras/sty2340}, \href {https://ui.adsabs.harvard.edu/abs/2018MNRAS.481.1075S} {481, 1075}

\bibitem[\protect\citeauthoryear{{Simonte}, {Vazza}, {Brighenti}, {Br{\"u}ggen}, {Jones}  \& {Angelinelli}}{{Simonte} et~al.}{2022}]{Simonte2022}
{Simonte} M.,  {Vazza} F.,  {Brighenti} F.,  {Br{\"u}ggen} M.,  {Jones} T.~W.,   {Angelinelli} M.,  2022, \mn@doi [\aap] {10.1051/0004-6361/202141703}, \href {https://ui.adsabs.harvard.edu/abs/2022A&A...658A.149S} {658, A149}

\bibitem[\protect\citeauthoryear{{Springel}}{{Springel}}{2010}]{Springel2010}
{Springel} V.,  2010, \mn@doi [\mnras] {10.1111/j.1365-2966.2009.15715.x}, \href {https://ui.adsabs.harvard.edu/abs/2010MNRAS.401..791S} {401, 791}

\bibitem[\protect\citeauthoryear{{Springel} et~al.,}{{Springel} et~al.}{2018}]{Springel2018}
{Springel} V.,  et~al., 2018, \mn@doi [\mnras] {10.1093/mnras/stx3304}, \href {https://ui.adsabs.harvard.edu/abs/2018MNRAS.475..676S} {475, 676}

\bibitem[\protect\citeauthoryear{{Subramanian}, {Shukurov}  \& {Haugen}}{{Subramanian} et~al.}{2006}]{Subramanian2006}
{Subramanian} K.,  {Shukurov} A.,   {Haugen} N. E.~L.,  2006, \mn@doi [\mnras] {10.1111/j.1365-2966.2006.09918.x}, \href {https://ui.adsabs.harvard.edu/abs/2006MNRAS.366.1437S} {366, 1437}

\bibitem[\protect\citeauthoryear{{Vacca}, {Murgia}, {Govoni}, {Feretti}, {Giovannini}, {Orr{\`u}}  \& {Bonafede}}{{Vacca} et~al.}{2010}]{Vacca2010}
{Vacca} V.,  {Murgia} M.,  {Govoni} F.,  {Feretti} L.,  {Giovannini} G.,  {Orr{\`u}} E.,   {Bonafede} A.,  2010, \mn@doi [\aap] {10.1051/0004-6361/200913060}, \href {https://ui.adsabs.harvard.edu/abs/2010A&A...514A..71V} {514, A71}

\bibitem[\protect\citeauthoryear{{Vacca}, {Murgia}, {Govoni}, {Feretti}, {Giovannini}, {Perley}  \& {Taylor}}{{Vacca} et~al.}{2012}]{Vacca2012}
{Vacca} V.,  {Murgia} M.,  {Govoni} F.,  {Feretti} L.,  {Giovannini} G.,  {Perley} R.~A.,   {Taylor} G.~B.,  2012, \mn@doi [\aap] {10.1051/0004-6361/201116622}, \href {https://ui.adsabs.harvard.edu/abs/2012A&A...540A..38V} {540, A38}

\bibitem[\protect\citeauthoryear{{Valdarnini}}{{Valdarnini}}{2011}]{Valdarnini2011}
{Valdarnini} R.,  2011, \mn@doi [\aap] {10.1051/0004-6361/201015340}, \href {https://ui.adsabs.harvard.edu/abs/2011A&A...526A.158V} {526, A158}

\bibitem[\protect\citeauthoryear{{Valdarnini}}{{Valdarnini}}{2019}]{Valdarnini2019}
{Valdarnini} R.,  2019, \mn@doi [\apj] {10.3847/1538-4357/ab0964}, \href {https://ui.adsabs.harvard.edu/abs/2019ApJ...874...42V} {874, 42}

\bibitem[\protect\citeauthoryear{{Vazza}, {Brunetti}, {Gheller}, {Brunino}  \& {Br{\"u}ggen}}{{Vazza} et~al.}{2011}]{Vazza2011}
{Vazza} F.,  {Brunetti} G.,  {Gheller} C.,  {Brunino} R.,   {Br{\"u}ggen} M.,  2011, \mn@doi [\aap] {10.1051/0004-6361/201016015}, \href {https://ui.adsabs.harvard.edu/abs/2011A&A...529A..17V} {529, A17}

\bibitem[\protect\citeauthoryear{{Vazza}, {Roediger}  \& {Br{\"u}ggen}}{{Vazza} et~al.}{2012}]{Vazza2012}
{Vazza} F.,  {Roediger} E.,   {Br{\"u}ggen} M.,  2012, \mn@doi [\aap] {10.1051/0004-6361/201118688}, \href {https://ui.adsabs.harvard.edu/abs/2012A&A...544A.103V} {544, A103}

\bibitem[\protect\citeauthoryear{{Vazza}, {Jones}, {Br{\"u}ggen}, {Brunetti}, {Gheller}, {Porter}  \& {Ryu}}{{Vazza} et~al.}{2017}]{Vazza2017}
{Vazza} F.,  {Jones} T.~W.,  {Br{\"u}ggen} M.,  {Brunetti} G.,  {Gheller} C.,  {Porter} D.,   {Ryu} D.,  2017, \mn@doi [\mnras] {10.1093/mnras/stw2351}, \href {https://ui.adsabs.harvard.edu/abs/2017MNRAS.464..210V} {464, 210}

\bibitem[\protect\citeauthoryear{{Vazza}, {Angelinelli}, {Jones}, {Eckert}, {Br{\"u}ggen}, {Brunetti}  \& {Gheller}}{{Vazza} et~al.}{2018}]{Vazza2018}
{Vazza} F.,  {Angelinelli} M.,  {Jones} T.~W.,  {Eckert} D.,  {Br{\"u}ggen} M.,  {Brunetti} G.,   {Gheller} C.,  2018, \mn@doi [\mnras] {10.1093/mnrasl/sly172}, \href {https://ui.adsabs.harvard.edu/abs/2018MNRAS.481L.120V} {481, L120}

\bibitem[\protect\citeauthoryear{{Vogt} \& {En{\ss}lin}}{{Vogt} \& {En{\ss}lin}}{2003}]{Vogt2003}
{Vogt} C.,  {En{\ss}lin} T.~A.,  2003, \mn@doi [\aap] {10.1051/0004-6361:20031434}, \href {https://ui.adsabs.harvard.edu/abs/2003A&A...412..373V} {412, 373}

\bibitem[\protect\citeauthoryear{{Vogt} \& {En{\ss}lin}}{{Vogt} \& {En{\ss}lin}}{2005}]{Vogt2005}
{Vogt} C.,  {En{\ss}lin} T.~A.,  2005, \mn@doi [\aap] {10.1051/0004-6361:20041839}, \href {https://ui.adsabs.harvard.edu/abs/2005A&A...434...67V} {434, 67}

\bibitem[\protect\citeauthoryear{{Walker}, {Sanders}  \& {Fabian}}{{Walker} et~al.}{2015}]{Walker2015}
{Walker} S.~A.,  {Sanders} J.~S.,   {Fabian} A.~C.,  2015, \mn@doi [\mnras] {10.1093/mnras/stv1929}, \href {https://ui.adsabs.harvard.edu/abs/2015MNRAS.453.3699W} {453, 3699}

\bibitem[\protect\citeauthoryear{{Wang} \& {He}}{{Wang} \& {He}}{2021}]{Wang2021}
{Wang} Y.,  {He} P.,  2021, \mn@doi [Communications in Theoretical Physics] {10.1088/1572-9494/ac10be}, \href {https://ui.adsabs.harvard.edu/abs/2021CoTPh..73i5402W} {73, 095402}

\bibitem[\protect\citeauthoryear{{Wang} \& {He}}{{Wang} \& {He}}{2022}]{Wang2022b}
{Wang} Y.,  {He} P.,  2022, \mn@doi [\apj] {10.3847/1538-4357/ac7a3d}, \href {https://ui.adsabs.harvard.edu/abs/2022ApJ...934..112W} {934, 112}

\bibitem[\protect\citeauthoryear{{Wang} \& {He}}{{Wang} \& {He}}{2023}]{Wang2023}
{Wang} Y.,  {He} P.,  2023, \mn@doi [RAS Techniques and Instruments] {10.1093/rasti/rzad020}, \href {https://ui.adsabs.harvard.edu/abs/2023RASTI...2..307W} {2, 307}

\bibitem[\protect\citeauthoryear{{Wang} \& {He}}{{Wang} \& {He}}{2024a}]{Wang2024b}
{Wang} Y.,  {He} P.,  2024a, \mn@doi [\apj, submitted, arXiv:2404.11255] {10.48550/arXiv.2404.11255}

\bibitem[\protect\citeauthoryear{{Wang} \& {He}}{{Wang} \& {He}}{2024b}]{Wang2024a}
{Wang} Y.,  {He} P.,  2024b, \mn@doi [\mnras] {10.1093/mnras/stae229}, \href {https://ui.adsabs.harvard.edu/abs/2024MNRAS.528.3797W} {528, 3797}

\bibitem[\protect\citeauthoryear{{Wang}, {Yang}  \& {He}}{{Wang} et~al.}{2022}]{Wang2022a}
{Wang} Y.,  {Yang} H.-Y.,   {He} P.,  2022, \mn@doi [\apj] {10.3847/1538-4357/ac752c}, \href {https://ui.adsabs.harvard.edu/abs/2022ApJ...934...77W} {934, 77}

\bibitem[\protect\citeauthoryear{{Wang}, {Oh}  \& {Ruszkowski}}{{Wang} et~al.}{2023}]{Wangc2023}
{Wang} C.,  {Oh} S.~P.,   {Ruszkowski} M.,  2023, \mn@doi [\mnras] {10.1093/mnras/stad003}, \href {https://ui.adsabs.harvard.edu/abs/2023MNRAS.519.4408W} {519, 4408}

\bibitem[\protect\citeauthoryear{{Zhu}, {Feng}  \& {Fang}}{{Zhu} et~al.}{2010}]{Zhu2010}
{Zhu} W.,  {Feng} L.-l.,   {Fang} L.-Z.,  2010, \mn@doi [\apj] {10.1088/0004-637X/712/1/1}, \href {https://ui.adsabs.harvard.edu/abs/2010ApJ...712....1Z} {712, 1}

\bibitem[\protect\citeauthoryear{{Zhu}, {Feng}  \& {Fang}}{{Zhu} et~al.}{2011}]{Zhu2011a}
{Zhu} W.,  {Feng} L.-L.,   {Fang} L.-Z.,  2011, \mn@doi [\mnras] {10.1111/j.1365-2966.2011.18640.x}, \href {https://ui.adsabs.harvard.edu/abs/2011MNRAS.415.1093Z} {415, 1093}

\bibitem[\protect\citeauthoryear{{Zhu}, {Feng}, {Xia}, {Shu}, {Gu}  \& {Fang}}{{Zhu} et~al.}{2013}]{Zhu2013}
{Zhu} W.,  {Feng} L.-l.,  {Xia} Y.,  {Shu} C.-W.,  {Gu} Q.,   {Fang} L.-Z.,  2013, \mn@doi [\apj] {10.1088/0004-637X/777/1/48}, \href {https://ui.adsabs.harvard.edu/abs/2013ApJ...777...48Z} {777, 48}

\bibitem[\protect\citeauthoryear{{Zhuravleva} et~al.,}{{Zhuravleva} et~al.}{2013}]{Zhuravleva2013}
{Zhuravleva} I.,  et~al., 2013, \mn@doi [\mnras] {10.1093/mnras/stt1506}, \href {https://ui.adsabs.harvard.edu/abs/2013MNRAS.435.3111Z} {435, 3111}

\bibitem[\protect\citeauthoryear{{Zhuravleva} et~al.,}{{Zhuravleva} et~al.}{2014}]{Zhuravleva2014}
{Zhuravleva} I.,  et~al., 2014, \mn@doi [\nat] {10.1038/nature13830}, \href {https://ui.adsabs.harvard.edu/abs/2014Natur.515...85Z} {515, 85}

\bibitem[\protect\citeauthoryear{{Zhuravleva}, {Allen}, {Mantz}  \& {Werner}}{{Zhuravleva} et~al.}{2018}]{Zhuravleva2018}
{Zhuravleva} I.,  {Allen} S.~W.,  {Mantz} A.,   {Werner} N.,  2018, \mn@doi [\apj] {10.3847/1538-4357/aadae3}, \href {https://ui.adsabs.harvard.edu/abs/2018ApJ...865...53Z} {865, 53}

\makeatother
\end{thebibliography}
\bibliographystyle{mnras}
\end{document}